\documentclass{WileyMSP-template}
\usepackage{amsmath}
\usepackage{xcolor}
\begin{document}

\pagestyle{fancy}
\rhead{\includegraphics[width=2.5cm]{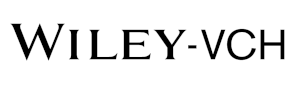}}

\title{Rotating fluorescent nanodiamond assemblies with focused Laguerre-Gaussian beams}

\maketitle


\author{Adam Stewart}
\author{Anthony J. El‐Helou}
\author{Ying Zhu}
\author{David McGloin}
\author{David A. Simpson}
\author{Peter J. Reece*}


\dedication{}

\begin{affiliations}

Adam Stewart, A/Prof. Peter J. Reece*\\
School of Physics, The University of New South Wales, Sydney, 2052, NSW, Australia\\
Email Address: p.reece@unsw.edu.au

Anthony J. El-Helou, Dr. Ying Zhu\\
School of Biomedical Engineering, Faculty of Engineering and IT, University of Technology Sydney, Sydney, 2007, NSW, Australia\\

Dr. Ying Zhu\\
School of Clinical Medicine, The University of New South Wales, Sydney, 2052, NSW, Australia\\

Prof. David McGloin\\
School of Natural and Computing Science, University of Aberdeen, Aberdeen, AB24 3FX, UK\\

Prof. David A. Simpson\\
Department of Physics, The University of Melbourne, Parkville, 3052, NSW, Australia\\

\end{affiliations}


\keywords{Optical Tweezers, Nitrogen-Vacancy Centers, Optically Detected Magnetic Resonance, Orbital Angular Momentum}

\begin{abstract}

Optical tweezers which utilize structured light fields enable the rotation of trapped nanoparticles through the transfer of orbital angular momentum (OAM) from holographically generated Laguerre-Gaussian (LG) modes. In this research we use OAM transfer to demonstrate controlled rotation of bright fluorescent nanodiamond clusters assembled in a focused higher-order LG beam. We find that the assemblies can be effectively rotated in a two-dimensional optical trap with orbital frequencies of up to 5 Hz. We use video tracking to explore the Brownian dynamics of such a trapping arrangement and look at the impact of orientation stability on measurements of optically detected magnetic resonance (ODMR) with an applied weak external magnetic field. By collecting ODMR spectra at multiple points along the orbit, we show that the constrained two-dimensional motion can provide additional insights for vector magnetic field reconstruction.

\end{abstract}


\section{Introduction}

Nanoscale vector magnetometry using nitrogen vacancy (NV\textsuperscript{-}) centre defects in diamond can provide precise information about the generation and evolution of sources of magnetic fields with high spatial resolution \cite{Doherty:2013}. An important factor in the determination of vector magnetic fields is the relative orientation of the NV defect's polar axis, as determined by the crystal lattice. In sensing based on diamond substrates, this information is known with a high degree of accuracy, however for sensing based on single nanodiamonds (NDs), the crystal can take on any possible orientation and needs to be considered as part of the reconstruction of the field. In sensing applications involving freely diffusing NDs, this is a dynamically varying parameter that, unaccounted for, can limit the accuracy and efficacy of the sensing platform \cite{McGuinness:2011aa}.

Optical tweezers, which have had some success in integration with NV based magnetic resonance sensing, provide a way of controlling the orientation of NDs in solution whilst still retaining its mobility within a sample volume. Specifically, shape-induced birefringence that occurs naturally in ball-milled diamond nanopowders can be used to control the alignment of a single optically trapped ND in a linearly polarised laser \cite{Geiselmann2013}. This has enabled the resolution of Zeeman-split energy levels from the four different NV orientations in the crystal lattice under an applied magnetic field, through the use of optically detected magnetic resonance (ODMR) \cite{Russell2021}. 

Whilst naturally occurring shapes can provide some degrees of confinement, significant stochastic forces still persist, and if the restoring torques are not sufficient, the NDs will suffer uncontrolled angular diffusion. This is typically the case for smaller NDs and those that are spherical in shape. Recently we showed that better stability and orientation control could be achieved using ND nano-structures formed by self-assembly inside the optical tweezers \cite{Stewart:2024}. These elongated structures that orient along the propagation axis exhibit significantly lower Brownian diffusion and can be controllably rotated using circularly polarised light. Despite the poly-crystallinity of the nano-assemblies, sensing at the resolution of a single nanodiamond could still be achieved, through the use of a confocal microscopy techniques.

The axial alignment of diamond nano-assemblies is favorable for certain applications, such as scanning probe microscopy \cite{Nakayama:2007aa}\cite{Fang:2025}, but other modes of optical manipulation, available using complex field shaping, can open new opportunities for nanoscale sensing. In particular for 2D trapping geometries, Laguerre-Gaussian (LG) beams, which support orbital angular momentum (OAM), can be used to rotate absorptive and birefringent particles\cite{He:1995}. This technique has also been applied at the nanoscale for rotating nanoparticles, such as gold nanospheres \cite{Dienerowitz:08,Shen:2019}. In addition, the mechanical equivalence of OAM and spin angular momentum (SAM)\cite{Simpson:97} transfer for absorptive particles has been explored using circularly polarised light \cite{Shao:2015}. In terms of potential advantages of quantum sensing, optically rotating ND sensors can in principle improve coherence times and sensitivity to external fields\cite{Barry:2020}. Experiments have shown DC magnetic field up-conversion in optical rotation of levitated nanodiamond\cite{Wood:2018,Jin:2024}, and mechanical rotation of bulk diamond\cite{Wood:2022}.

In this paper, we explore the application of LG beams to two-dimensional optical manipulation of self-assembled ND structures, and its integration with magnetic resonance sensing. We show that controlled and repeatable rotation can be achieved for elongated nano-assemblies trapped against the top surface of a fluidic chamber using a high-order LG beam. Using video tracking, we resolve the position and orientation of the nano-assembly at each point on the elliptical path. By applying an excitation probe beam at a particular point on the ring, we are able to accumulate ODMR spectra over a number of cycles. We show that the angular diffusion of the orbiting particles under small magnetic fields can be accounted for in the observed ODMR spectral features for this particular geometry. By collecting ODMR spectra at multiple points along the orbit, we further show that the response is symmetric under half-orbit rotations about an in-plane magnetic field, suggesting minimal rotation about the long axis. These results indicate that 2D rotations provide a pathway to vector reconstruction of external magnetic fields and crystal orientation.

\section{Results and Discussion}

\begin{figure}[]
\centering\includegraphics[width=\textwidth]{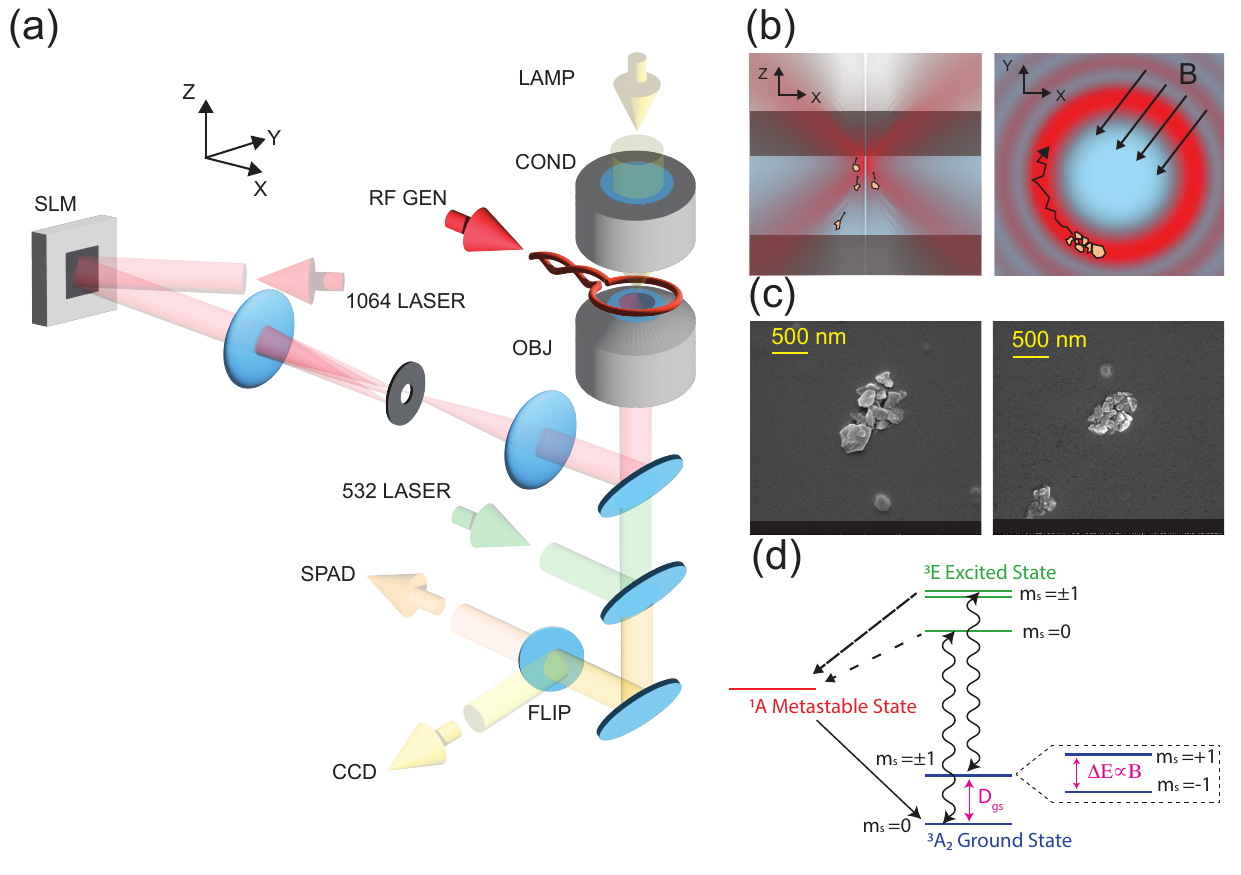}
\caption{(a) Experimental apparatus: an infrared (1064 nm) laser reflected from a spatial light modulator (SLM) displaying an appropriate phase-mask is used to create a Laguerre-Gaussian mode at the focus of a high magnification objective (OBJ). Diascopic bright field illumination is used with a CCD camera to observe the motion of the nanodiamond (ND) assemblies. A co-linearly aligned green (532 nm) laser is used to excite fluorescence from the NV\textsuperscript{-} defect centres, which is collected with a single photon avalanche diode (SPAD) via a flipper (FLIP) mirror. A 2-4 GHz RF signal applied to the sample via a small loop antenna is used to drive the NV\textsuperscript{-} electronic spin states which creates optical contrast in the fluorescence intensity. (b) In a typical experiment, freely dispersed NDs are pushed to the top cover-slip where they accumulated into aggregates in the high intensity ring. The ND assemblies begin to orbit with controlled rotation due to the transfer of optical angular momentum. The orientation of the assemblies changes with respect to an external magnetic field. (c) Example scanning electron microscope (SEM) images of ND assemblies formed inside the optical tweezers apparatus, which influences the optically detected magnetic resonance (ODMR). (d) ODMR contrast is produced between the $m_s = 0$ and $m_s = \pm 1$ by a preferential relaxation from triplet excited (\textsuperscript{3}E) to ground state (\textsuperscript{3}A) via the non-radiative singlet state. The application of a magnetic field will create a Zeeman splitting of the $m_s = \pm 1$ states.}\label{overview}
\end{figure}

Two-dimensional (2D) optical trapping experiments were performed at the top interface of a thin (\(\simeq\)150 \(\mu\)m) microfluidic chamber in a custom-built inverted microscope. Details of the arrangement can be found in Section \ref{sec:experimental} and in Fig.\ref{overview}(a).

The initial phase of accumulation and trapping of NDs at the top interface of the chamber is similar for all beam profiles surveyed Fig.\ref{overview}(b). Starting with NDs freely dispersed within the chamber, when the trap is initiated, the particles are driven to the surface by a combination of axial radiation pressure and lateral gradient forces. As the NDs are predominantly transparent at the trapping wavelength, we proposel that optical forces are dominant. Confined particles initially exhibit strong random motion within the trap. For LG modes, the onset of clear orbital motion is associated with the nucleation of one or more clusters. Depending on the particular experiment, the clusters may remain separate, but show correlated motion, or will form into larger scale assemblies. Examples of assembled structures adhered to the coverslip and imaged using scanning electron microscopy (SEM) are presented in Fig.\ref{overview}(b). Clustering of multiple NDs into elongated shapes can be seen, and in some cases, there appear to be multiple clumps. Qualitatively we find that the orbital motion of the clusters are moderated by the order of the LG beam as well as the incident power. Lower order modes and higher powers exhibit the fastest tangential velocities, however there is a trade off between angular trapping stability and regular orbital motion. An example of particle motion for different LG modes is provided in the supplementary videos (Fig.\,S1). In addition, changing the focus of the beam relative to the coverslip can also be used to moderate the particle motion. For this study, we focus our attention on the $l = 7$ mode which gives optimal particle motion for the available optical powers. Videos showing the power dependence of the optical rotation is demonstrated in Fig.\,S2.   

\begin{figure}[]
\centering\includegraphics[width=\textwidth]{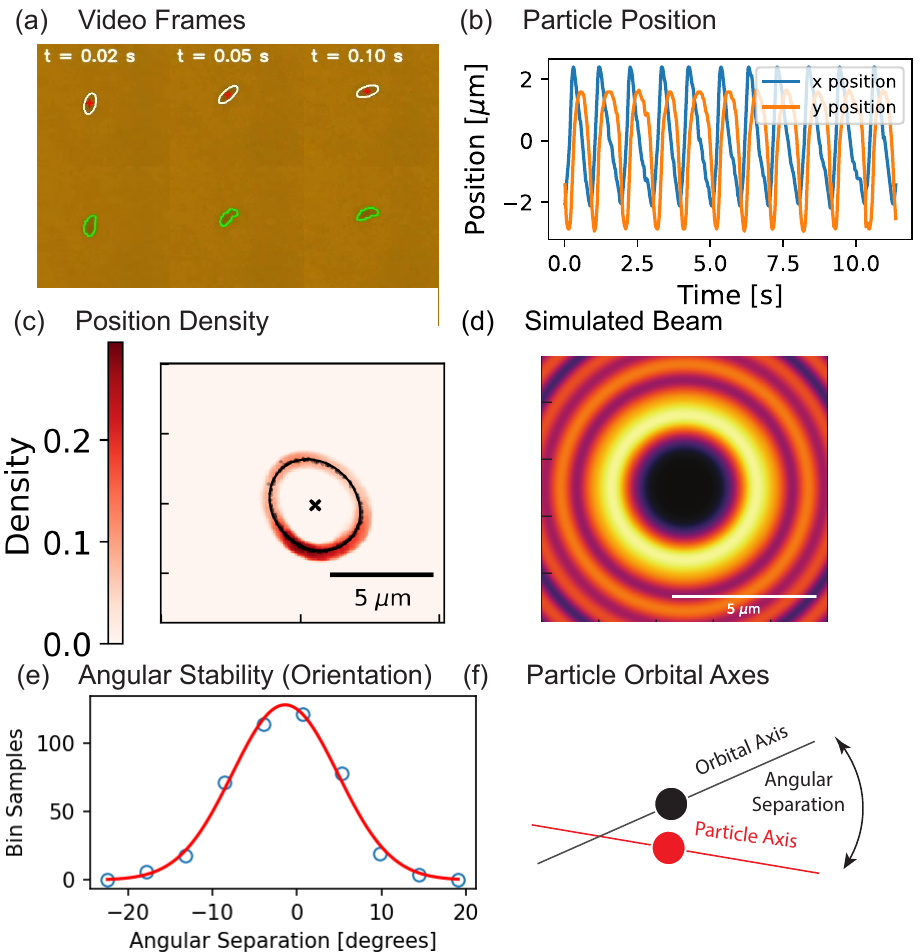}
\caption{(a) CCD camera frames with the tracked particle boundary (green), center of mass position (red) and elliptical fit (white). (b) The particles position over time as it orbits around the center of the beam. (c) The position density function of the contour fit over the measurement. The black line is the elliptical fit and the cross marks the center. (d) A simulated Laguerre-Gaussian beam profile of the optical tweezers, with azimuthal mode $l = +7$ and radial mode $m = 0$ (TEM$_{07}$), 5 $\mu$m displacement in +z from focus. (e) A histogram of the angular difference between the orientation of the rotating particle and the tangent of the nearest position on the elliptical axis, for all the frames in the dataset. The fit (red) to the data (blue) has mean value $|\mu|$ = 1.4 $\deg$ and standard deviation $\sigma$ = 6.3 $\deg$. (f) A graphic of the angular separation and the particle orientation.}\label{Trap}
\end{figure}

In Fig.\,\ref{Trap} we present an analysis of the dynamics of a particular assembly of NDs in an LG beam with azimuthal mode of +7. For this data set, images were recorded at a frame rate of 40\,Hz for 11\,s (exposure time = 24\,ms). For each frame we extract the central position of the clusters as well as the orientation of the cluster as defined by the major axis of an ellipse fitted to the contour of the object. Example frames showing the orbiting NDs with accompanying position and orientation information is presented in Fig.\,\ref{Trap}(a), along with the time series traces of the x and y positions in Fig.\,\ref{Trap}(b). The motion is observed to be regular and traces an elliptical path with an orbital period of approximately 1\,s. The relatively slow orbital motion is chosen in this instance to minimize image blur on the video recordings, however it is possible to reach orbital frequencies of up to 5\,Hz at higher powers. A plot of the position density for this trajectory is presented in Fig.\,\ref{Trap}(c) together with a simulation of the expected beam profile in Fig.\,\ref{Trap}(d), which predicts a comparable orbit. It is interesting to note that, during assembly, particles can initially accumulate in the higher order rings. From the plot we see that the orbital motion is not uniform and this may reflect hot spots in the beam profile. For applications such as vector magnetometry, the ability to maintain a fixed orientation relative the external field will directly impact sensitivity limits. In order to quantify this for our orbiting nanoassemblies, we define an angular stability in terms of the deviation from an orientation tangential to the elliptical orbit that is defined by the centroid motion (see. Fig.\,\ref{Trap}(e)). The results of this analysis are presented in Fig\,\ref{Trap}(d). Notably, we observe a fluctuation in the orientation of approximately $13^{\circ}$ averaged over the entire orbit.

To understand the impact of angular stability on magnetic field sensing, we consider the expected broadening of spectral features due to corresponding changes in the orientation of the NV axis relative to an external field. The Hamiltonian of the $^3A_2$ ground state in an external magnetic field B can be written $H=DS_z^2 + 2E(S_x^2-S_y^2) + \gamma_{e}\vec{B}\cdot\vec{S}$, where B is static field vector, $\vec{S}$ is the NV spin matrix, $D$ the zero-field resonance, $E$ the strain, and $\gamma_{e}$ is the electron gyromagnetic ratio ($\gamma_e \approx$ 28.0 MHz/mT). A small change in angle $\Delta\theta$ of B from the NV axis will cause the Zeeman energy to change by $\Delta E$. The uncertainty in orientation can be related to the change in resonance energy by a first-order perturbation $\Delta E$ about the orientation $\Delta\theta$ written as

\begin{equation}
    \Delta E / |\vec{B}|= \Delta \theta  \gamma_e  \sin \theta
\end{equation}

where $\theta = \pi/4$. The maximal uncertainty in resonance energy occurs when the magnetic field makes an angle $\pi/4$ to one of the NV axes. In Fig.\,\ref{Trap}, our observed 2$\sigma$ angular stability was $\Delta\theta = \pm13^\circ$ in an external field of $\approx$ 1.3 mT. The maximum change in resonance frequency due to angular dispersion in that case is $\Delta E / |B| = \pm4.6$ MHz/mT. This represents the most significant broadening expected for this particular trapping configuration. Any other NV orientation should be less sensitive to angular diffusion. This level of broadening is comparable to the natural line-widths observed in bright fluorescent NDs with no magnetic field applied, and would represent significant broadening at higher fields.

\begin{figure}[]
\centering\includegraphics[width=\textwidth]{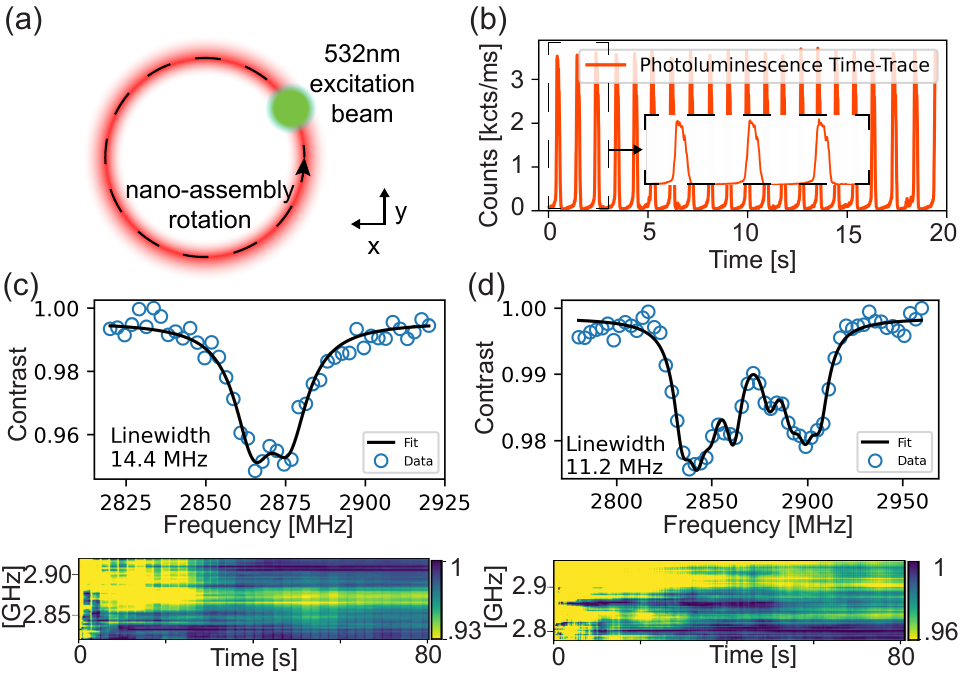}
\caption{(a) The nanodiamond assemblies orbit around the beam axis (dotted line) and are excited with a Gaussian mode solid-state green laser.  The excitation beam is positioned at a point on the orbital path and stimulates emission of the nitrogen-vacancy (NV-) centers (b) The NV- center fluorescence is collected with an avalanche photodiode detector (APD). The time-trace shows periodic peaks which occur when the nanodiamonds coincide with the excitation beam. The dotted region is zoomed to show the shape of individual peaks. (c) The optically detected magnetic resonance (ODMR) spectrum in ambient magnetic field condition. The fit (black line) has a spectral doublet Lorentzian line-shape with center frequency 2870 MHz, separation of 11.0 $\pm$ 0.6 MHz, and linewidth of 14.4 $\pm$ 1.4 MHz. The panel underneath the spectrum contains the cumulative average, which illustrates the
convergence of repeated ODMR sweeps over the measurement. The cumulative average is calculated by sequentially averaging spectra collected during rotation. (d) The ODMR spectrum (and cumulative average panel below) when a rare-earth magnet is positioned to create an external magnetic field $\approx$ 1.3 mT. The fit has a 8-peak characteristic Lorentzian shape, where the linewidth is 11.2 $\pm$ 0.9 MHz.}\label{ODMR}
\end{figure}

To test the efficacy of this trapping geometry for magnetic resonance sensing, a green laser was used to excite fluorescence from the ND assembly at a particular point on the orbital path, as depicted in Fig.\,\ref{ODMR}(a). An example time series trace of the resultant fluorescence intensity is plotted in Fig.\,\ref{ODMR}(b), which appears as a series of peaks where the orbiting nano-assembly transits the excitation spot. The inset shows that the peak profiles have a consistent shape, suggesting that the orientation of the objects remain constant. Using a time average acquisition, the ODMR spectrum was successfully recorded both in the absence and presence of background magnetic field. In Fig.\,\ref{ODMR}(c) the average ODMR spectrum is plotted along with a 2D plot showing the evolution of the cumulative average. We see that a clear characteristic double Lorentzian peak appears within 20 seconds of acquisition, with a line width comparable to those expected from commercial bright fluorescent NDs. Similarly, with an external magnetic field applied (B = 1.3 mT) to the sample, we observe peak splitting with a corresponding reduction of optical contrast Fig.\,\ref{ODMR}(d). Of note, the fitted spectral widths of these peaks are comparable to zero B-field case, suggesting that orientational effects are not a limiting factor in this case. We calculate the shot-noise limited DC magnetic field sensitivity\cite{Barry:2020} to be $40.0$ $\pm$ 7.8 $ \mu\text{T}$/$\sqrt{\text{Hz}}$, from the photon count rate 450 $\pm$ 4 kcts/s, contrast 0.011 $\pm$ 0.002 and linewidth 11.2 $\pm$ 0.9 MHz for the ODMR spectrum in Fig.\,\ref{ODMR}(d).

\begin{figure}[]
\centering\includegraphics[width=
\textwidth]{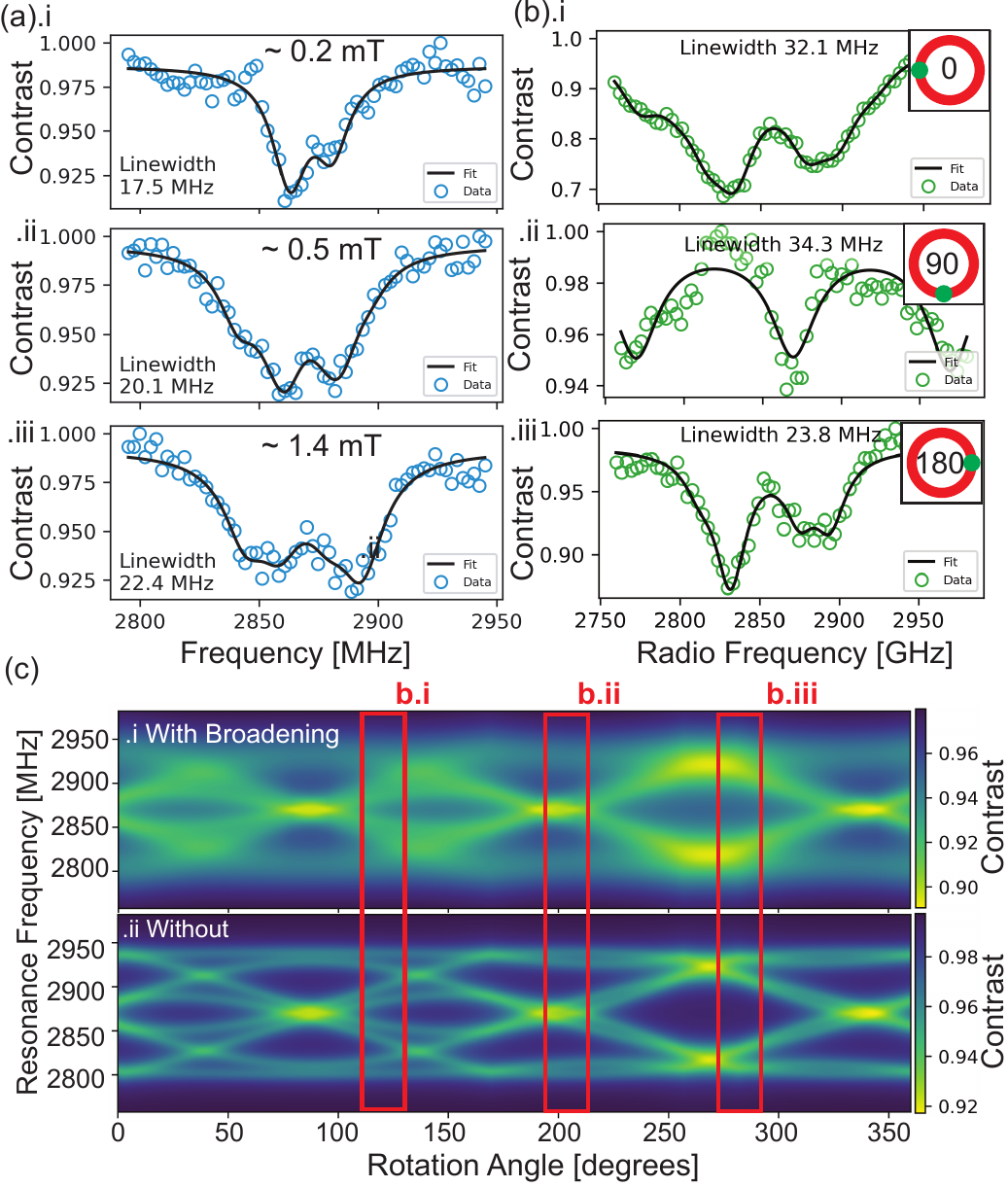}
\caption{(a).i-.iii. Optically Detected Magnetic Resonance (ODMR) spectra of rotating nanodiamond assemblies measured under increasing external magnetic field strengths in a Laguerre-Gaussian TEM$_{07}$ beam. (b).i-.iii ODMR measurements collected with the excitation focused at orbital positions corresponding to a relative yaw angle of $0^{\circ}$, $90^{\circ}$ and $180^{\circ}$. The inset image in the top right depicts the orbital position where the ODMR spectra were collected. (c).i-.ii. Simulated ODMR spectra for orbital rotations around the beam center in an external magnetic field which is matched to our system. The locations highlighted are separated by $90^{\circ}$ and matched to the observed spectra in (b).i-.iii.}\label{ODMR2}
\end{figure}

Whilst the presented results don't appear to be angular diffusion limited, one may consider circumstances where the orientation fluctuations may begin to dominate the linewidth of the ODMR spectrum. For example, if the trapping power of the infrared laser were reduced, the decrease in angular stability should also lead to a similar broadening. For the second case, the nonlinear response of the trapping geometry, makes it difficult to achieve controlled rotation at these lower powers. Alternatively, if the applied external magnetic field increases, then the corresponding angular dependent peak shift would be sufficiently large to broaden the spectral features. In Fig.\,\ref{ODMR2}(a).i-.iii we present a set of results showing ODMR spectra from an orbiting ND assembly at different applied magnetic field strengths. At low fields (0.2 mT) we observe a narrow spectral doublet, that is a combination of Zeeman splitting and intrinsic effects. With increasing magnetic field (0.5 mT and 1.4 mT) the peaks split further with an increase in linewidth. The composition of the peaks include a convolution of different splittings due to NV axis projections as well as broadening due to angular diffusion.

Extending on the single point photoluminescence measurements presented, we may also probe multiple points along the particle trajectory. For this geometry, and in the presence of a uniform magnetic field, orbiting about the LG beam modulates the orientation of the NV ensemble relative to the external field, and hence the Zeeman splitting energies. For in-plane, or near in-plane external magnetic fields, the ODMR spectrum can be symmetric for angular rotations around the beam axis, assuming other rotations are fixed. Re-construction of the NV axes in the nano-assembly orientation can be achieved in this case, for example, by probing the ODMR spectrum at different points on the orbit and comparing to simulated values. We performed experiments to demonstrate this effect, measuring the ODMR spectrum at points on the orbit corresponding to relative orientation changes of $\sim90^{\circ}$ in the presence of a magnetic field. In Fig.\,\ref{ODMR2}(b)i.-iii. we present three measurements of the ODMR spectrum taken points on the beam orbit indicated in the inset. The ODMR spectra at 0$^{\circ}$ and 180$^{\circ}$ are spectral doublets and remain qualitatively similar in frequency and shape, which we attribute to the symmetry of the arrangement, and a lack of orientation change in the pitch and roll directions. At 90$^{\circ}$ a spectral triplet is observed which strongly suggests a change in projection of the external field to the defect centers.

This effect can be further explored in simulations. In  Fig.\,\ref{ODMR2}(c)i.-ii, we present simulations where the ODMR spectrum is calculated for positions on a circular orbit with a fixed pitch and roll. We use a technique similar to that found in \cite{Feng:2021}, where Euler angles are used to define the orientation of the external magnetic field and the ensemble of NV centers inside the crystal. We calculate the resonance energies from the first-order Zeeman effect which is the result of a projection of the external field onto the NV axes. The ODMR spectrum is obtained through summation of Lorentzian functions at the resonance energies. The orbital motion of the particle is then simulated by sweeping the Euler angle one full rotation, which allows us to obtain the rotation angle dependence on the ODMR spectrum. We can simulate the broadening due to angular diffusion by perturbing the linewidth $\Gamma \rightarrow \Gamma + \Delta\Gamma$, where $\Delta\Gamma = \Gamma\cdot|\omega_{\pm} - D|\cdot\Delta\text{E}/|B|$ and $\omega_{\pm}$ are the energy eigenvalues, and $D$ is the zero-field splitting. In these conditions, the defect axes with larger Zeeman energies also have a larger linewidth. A comparison with and without angular broadening for the same angle sweep is presented in Fig.\,\ref{ODMR2}(c)i.-ii. The result of this is broader spectral lines, which might limit the accuracy. By inspection, we find possible candidate orientations for the measured spectra, which are indicated as red boxes. Based on these projections, we estimate that the external magnetic field strength is $\approx$ 2.4 mT, and orientation in cartesian coordinates [1.48, 1.89, 0.03] mT (38$^\circ$ to the yaw axis).

In moving to a 2D trapping geometry we have lost some of the flexibility afforded to the gradient force optical tweezers. However, we have gained considerable insights into the dynamics of the trapped nano-assemblies through video tracking – information that would otherwise have been hidden. Further, the uniform circular motion provides us with a constrained trajectory that can be used to more readily discern the applied magnetic field orientation. One may also consider the use of other complex beam shaping arrangements for controlling the motion and orientation, however the Laguerre-Gauss modes do provide a particularly simple mapping between position and orientation. Once the orientation of the ND probe has been established, it should be possible to interrogate local heterogenous fields from unknown sources near the surface. Such experiments could be aided using nano-engineered shapes that are optimized for stability, crystallinity and electronic properties, however the facile nature of the self-assembly process and the ability to provide in situ characterization of the magnetic sensing properties is an enabling feature of this current approach.

\section{Conclusion}

In conclusion we have explored a new mode of optical manipulation for self-assembled nanodiamonds that incorporates orbital angular momentum transfer. We have shown that elongated structures exhibit controlled orbiting in a higher order LG beam at rates of 1-5 Hz with a well defined orientation. We adapt our magnetic resonance sensing measurement protocol to successfully measure ODMR from the orbiting particle, and investigate the effect of orientation stability on the spectral response of the nitrogen-vacancy centers under an applied field. These findings enhance the prospects of future applications of vector magnetometery with trapped nanodiamonds.  

\section{Experimental Section}\label{sec:experimental}

For trapping, a 1064\,nm wavelength diode pumped solid-state laser (IR Ventus, Laser Quantum) is focused, via an imaging system, into the sample chamber through an oil-immersion objective (CFI Plan Fluor 100x  1.3 NA, Nikon). A spatial light modulator (SLM, Hamamatsu, LCOS-SLM x10468-03) placed at a plane conjugate to the objective back aperture is used to generate Laguerre-Gaussian modes with a holographic phase mask \cite{Gerchberg1972}. This study focused primarily on the use of an azimuthal mode of $l = 7$. A two-axis acousto-optic deflector (2D-AOD, Gooch \& Housego, 45035 AOBD), placed at a second conjugate back aperture plane, is used to control the power of the trapping laser and also to modulate the trap during magnetic resonance measurements.

The samples consist of commercially sourced bright fluorescent NDs (FND Biotech, Taiwan) that are dispersed in 5\,$\mu$L aliquots, prepared by diluting 1 part  stock concentration ($1$\,mg/mL) with 1000 parts of milliQ water. The NDs are high pressure high temperature (HPHT) processed and have a nominal effective diameter of 100\,nm, but are poly-dispersed in size and shape. These NDs are specifically prepared using a combination of ion irradiation and annealing to promote the formation of NV\textsuperscript{-} defects ($< 1000$ per particle)\cite{Hsiao2022, Hsiao2022b}. The aliquots are dropped on to microscope coverslips which had been surface treated with an air plasma (Handheld Corona Surface Treater, Aurora Pro Scientific), to prevent adhesion of the nanodiamonds to the trapping interface. During the experiments, individual NDs are accumulated slowly in the trap over several minutes. Assembled-nanostructures form into elongated rod-like structures via Van der Waals interactions. 

Nano-assembly motion is monitored under bright-field diascopic illumination and recorded using a CCD camera (Allied Vision Technologies, Stingray F145-C). The trapped assemblies are observed as a dark region where the particles have accumulated. Video recordings are processed with routines from an open-source computer vision library\cite{opencv_library}, and a video algorithm tracks the position of the rotating assemblies. A threshold and morphology step cleans the frame, and standard density fit for the contour of the particle position. The central moment is calculated from the contour to measure position. An elliptical fit is used to track the rotation of the major and minor axes. The orientation of the particle relative to the orbital path center measures the angular separation.

Optical excitation is performed with a 532 nm optically pumped semiconductor laser  (Verdi G7, Coherent) modulated with an acousto-optic modulator (AOM, Gooch \& Housego, 3110-120). The mechanism for ODMR contrast is presented in Fig.\,1(d), optical excitation excites electrons from the NV $^3\text{A}$ ground state to the $^3\text{E}$ excited state which then either relax back to the ground state via emission of a photon, or non-radiatively via singlet metastable states $^1\text{A}$ and $^1\text{E}$. The resonance energy between the $m_s=0$ and $m_s=\pm1$ is largely dependent on the presence of external magnetic fields via the Zeeman effect and strain in the crystal. A second SLM (Hamamatsu, LCOS-SLM x10468-01) and relay optics are used to align the excitation laser to points within the trapping region. Phonon-broadened fluorescence from the NV\textsuperscript{-} defects emission is collected through an optical band-pass filter (Semrock FF02-675/67-25) using a single-photon avalanche diode (SPAD, PicoQuant, $\tau$-SPAD). Photon counts are binned in 5 ms time intervals with a data acquisition (DAQ) card (National Instruments, NI PCI-6251) to create a fluorescence time trace. Pulsed ODMR measurements are performed by alternatively applying a 300 ns excitation laser pulse followed by a microwave pulse of 400\,ns via a low Q-factor loop antenna placed next to the sample surface. The timing for excitation and detection is controlled with pulse-sequences from a TTL generator (PulseBlasterESR-PRO-II). The microwaves are generated using a vector signal generator (Rohde \& Schwarz, SMBV100A) and power amplifier (Mini Circuits, ZHL-25W-63+). The resulting normalized fluorescence count rate (i.e. optical contrast) is recorded as a function of applied microwave frequency over the range of 2.7 to 3.0\,GHz. To observe the magnetic field dependence of the ODMR from the NDs in the trap, a small rare-earth permanent magnet is brought in close proximity to the trapping region. The average field at the trapping site is measured with a gaussmeter (Lake Shore Model 410) and transverse Hall probe (Lake Shore MST-9P04-410).

\medskip
\textbf{Supporting Information} \par 
Supporting Information is available from the Wiley Online Library or from the author.


\medskip

%
\bibliographystyle{MSP}
\bibliography{Rotating}

\begin{figure}
\textbf{Table of Contents}\\
\medskip
  \includegraphics{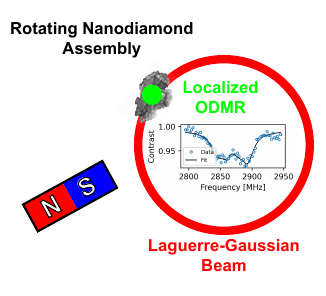}
  \medskip
  \caption*{Self-assembled fluorescent nanodiamond clusters are optically trapped and driven into controlled two-dimensional rotation with Laguerre–Gaussian beams. With localized optical excitation, optically detected magnetic resonance spectra are collected at defined points along the orbit in a uniform external magnetic field. This approach reveals orientation stability during rotation and establishes a route toward vector magnetic field reconstruction.}
\end{figure}

\end{document}